\documentclass[%
showpacs,
 amsmath,amssymb,
 aps,
 10pt,
twocolumn
]{revtex4-1}

\usepackage[usenames,dvipsnames]{color}
\usepackage{graphicx,textcase} 
\usepackage{dcolumn,overpic} 
\usepackage{bm,color}
\usepackage[T1]{fontenc}
\usepackage{ae,aecompl}
\usepackage{chngcntr}
\usepackage[para]{threeparttable}
\usepackage{etoolbox}
\usepackage{lipsum}
\AtEndEnvironment{table*}{\vskip10pt}{}{}

\begin{document}

\title{Layer-resolved magnetic exchange interactions of surfaces of late 3d elements: effects of electronic correlations}

\author{S. Keshavarz$^1$, Y.O. Kvashnin$^1$, I. Di Marco$^1$, A. Delin$^{1,2}$, M. I. Katsnelson$^{3,4}$, A. I. Lichtenstein$^{4,5}$, O. Eriksson$^1$}

\affiliation{$^1$Uppsala University, Department of Physics and Astronomy, Division of Materials Theory, Box 516, SE-751 20 Uppsala, Sweden}

\affiliation{$^2$KTH Royal Institute of Technology, Department of Materials and Nano Physics, Electrum 229, SE-164 40 Kista, Sweden}

\affiliation{$^3$Radboud University of Nijmegen, Institute for Molecules and Materials, Heijendaalseweg 135, 6525 AJ Nijmegen, The Netherlands}

\affiliation{$^4$Theoretical Physics and Applied Mathematics Department, Ural Federal University, Mira Street 19, 620002 Ekaterinburg, Russia}

\affiliation{$^5$Institute of Theoretical Physics, University of Hamburg, Jungiusstrasse 9, 20355 Hamburg, Germany}

\date{\today}

\begin{abstract}
We present the results of an \textit{ab initio} study of magnetic properties of Fe, Co and Ni surfaces. In particular, we discuss their electronic structure and magnetic exchange interactions ($J_{ij}$), as obtained by means of a combination of density functional theory and dynamical mean-field theory. 
All studied systems have a pronounced tendency to ferromagnetism both for bulk and surface atoms. 
The presence of narrow-band surface states is shown to enhance the magnetic moment as well as the exchange couplings.
The most interesting results were obtained for the Fe surface where the atoms have a tendency to couple antiferromagnetically with each other. 
This interaction is relatively small, when compared to interlayer ferromagnetic interaction, and strongly depends on the lattice parameter.
Local correlation effects are shown to lead to strong changes of the overall shape of the spectral functions.
However, they seem to not play a decisive role on the overall picture of the magnetic couplings studied here. We have also investigated the influence of correlations on the spin and orbital moments of the bulk-like and surface atoms. We found that dynamical correlations in general lead to enhanced values of the orbital moment. 

\end{abstract}

\maketitle

\section{Introduction}
Bulk Fe, Co and Ni are all classical examples of ferromagnets. 
However, when confined to two dimensions, these transition metals (TM) show a large panorama of fascinating magnetic properties and phenomena~\cite{Wiesendanger-STM,FeCo-superlatt-exp,Co-W001-Anders}.
For instance, thin layers of these atoms may show antiferromagnetic (AFM) behavior or even non-collinear spin structures, depending on the film thickness and/or the substrate~\cite{chumakov-FeW-noncoll,FeW-STM,Fe-Rh001-exp}.
The latter is known to play an important role, producing strains due to lattice mismatch and hybridizing with the TM states~\cite{TM-on-W-Blugel,FeW-Mills-2003,Fe-Rh100-Blugel,kudrnov}.
All above-mentioned effects contribute to the magnetic exchange interactions ($J_{ij}$), that are the relevant parameters of an effective spin-Hamiltonian which determine the Curie temperature and magnon dispersion of the material.
The latter two quantities are of particular importance for technological applications in, e. g., spintronic memory and logic devices. Hence, a fundamental understanding of the magnetic properties of these systems is needed.

The magnetism of surfaces has been of interest for quite some time. Initial studies were mainly focused on differences between surfaces and bulk properties. For instance, experimentally  it was for some time discussed that fcc Ni layers on top of a Cu substrate produced magnetically 'dead' layers, with an absence of magnetic moments~\cite{dead}. However subsequent experiments~\cite{kortright} and theory~\cite{ding,jepsen, wimmer} suggested that the spin-moments at surfaces in general are enhanced, since the bands are narrower. Later on relativistic electronic structure theory could analyze also the orbital moments of surfaces, and here the enhancement of the surface magnetism was found to be even larger than the spin-contribution for bcc Fe, hcp Co and fcc Ni~\cite{bruno, eriksson1, eriksson2}. These theoretical predictions were confirmed by experiments using x-ray magnetic circular dichroism~\cite{tischer}. 

Computational modeling is very important to investigate magnetic properties, as it gives a material specific description and makes it possible to disentangle all the relevant contributions.
Density functional theory (DFT) and its formal extensions~\cite{kohn,rung,mermin} give an excellent parameter-free description of ground-state properties of magnetic metals, including bulk structures as well as systems without three-dimensional periodicity such as surfaces, interfaces, thin films, disordered alloys and nanoparticles. However, several studies have emphasized the importance of including strong correlation effects in the electronic structure of bulk Fe, Co, and Ni. For instance, noncoherent features such as Hubbard bands and satellites which appear in the excitation spectra of the photoemission experiments~\cite{chan, guillot} can not be described by LDA and/or GGA. In addition, LDA calculations predicts too wide majority spin $3d$ band and overestimate the spin splitting for these materials~\cite{u1,u2,igor,orbital1}.

Correlation effects in transition metals are expected to be even more pronounced for the surface atoms, due to narrower bands and  reduced coordination numbers. In this article we report on a computational study of surface magnetism of TM slabs. The main focus is on the calculations of magnetic moments and inter-atomic exchange interactions ($J_{ij}$). The simulations are based on a combination of DFT and dynamical mean-field theory (DMFT). This technique, which is usually addressed as LDA+DMFT, is at the moment the state-of-the-art method to study strong correlations in materials at finite temperature~\cite{dmft1,dmft2,dmft3}. 

To the best of our knowledge, LDA+DMFT has not been previously applied to the exchange interactions of the transition metals surfaces. Even for standard DFT such simulations are rare, as most attention was focused on thin films on various substrates~\cite{turek, chuang, meng, deak, kudrnov}. One of the reasons for this is that many softwares are still based on the atomic-sphere approximation (ASA), which limits their use for studying surfaces, or low dimensional systems in general. The methods used in the present work do not suffer of this limitation.

This paper is organized as follows. In section \ref{theory} we briefly explain the computational scheme used in this work, as well as the implementation of the formalism described in Ref~\cite{liechten1} for evaluating the exchange parameters. The result of our electronic structure calculations and exchange interactions for Fe, Co and Ni surfaces are presented in section \ref{results}. The following section reports our investigation of the orbital polarizations for each slab. Finally, we draw our conclusions which will be followed by three Appendices. Appendix \ref{u} concerns the influence of the Hubbard U value, Appendix \ref{csc} is about the effect of full self-consistency over the charge density on the exchange parameters and Appendix \ref{qw} describes renormalization factors due to many-body effects.

\section{Theory}
\label{theory}

The electronic structure as well as the magnetic properties of the TM slabs were investigated in the framework of scalar-relativistic full-potential linear muffin-tin orbital (FP-LMTO) code RSPt~\cite{rspt}. Due to the full potential character the code does not have limitations dictated by the geometry of the problem under consideration. Moreover, due to the small number of basis functions, RSPt is particularly suitable for LDA+DMFT simulations with full self-consistence over self-energy and electron density. Details of this implementation were presented elsewhere~\cite{rspt,igor,patrik,oscar} and will not be repeated here. We redirect the reader to those references for a detailed overview of our formalism. 

Once the electronic structure was converged, the magnetic excitations were mapped onto the Heisenberg Hamiltonian:

\begin{eqnarray}
\hat H = -\sum_{i \ne j} J_{ij} \vec e_i \vec e_j ,
\end{eqnarray}
where $J_{ij}$ is an exchange interaction between the spins, located at sites $i$ and $j$, and $\vec e_i$ is a unity vector along the magnetization direction at the corresponding site.
We extracted the pair-wise exchange interactions by employing the method of infinitesimal rotation of the spins.
The exchange parameters were computed using the local magnetic force approach~\cite{liechten1, liechten2}, which reads:
\begin{equation}
\label{eq:jij}
J_{ij}=\frac{1}{4}Tr_{\omega,L}\left [ \hat\Sigma_i^s (i\omega_n) G_{ij}^\uparrow(i\omega_n) \hat\Sigma_j^s(i\omega_n) G_{ji}^\downarrow(i\omega_n)\right ],
\end{equation}
where $G^{\sigma}_{ij}$ is the inter-site Green's function, $\sigma$ denotes spin projection ($\sigma=\{\uparrow,\downarrow\}$), and the trace is taken over the fermionic Matsubara frequencies $i\omega_n$ and  the states characterised by an angular momentum quantum number $L$. The crucial quantity in Eq.~\ref{eq:jij} is the dynamical on-site exchange potential: 
\begin{equation}
\label{eq:dynamic_exch}
\hat\Sigma_i^s(i\omega_n) \equiv \left ( \hat{H}_i^\uparrow - \hat{H}_i^\downarrow \right )+\left ( \hat\Sigma_i^\uparrow (i\omega_n)-\hat\Sigma_i^\downarrow(i\omega_n)\right ) ,
\end{equation}
where $\hat{H}_i^{\sigma}$ is the local Hamiltonian matrix, obtained by solving the DFT equations and $\Sigma_i^\sigma$ is the self-energy, describing the electronic correlations. The self-energy appears only for LDA+DMFT calculations, and also enters the expression of the Green's function as:
\begin{equation}
\label{eq:gf}
\hat G_{ij}(i\omega_n)=\left \langle i\left |\frac{1}{i\omega_n-\hat{H}-\hat{\Sigma}(i\omega_n)}  \right |j \right \rangle .
\end{equation}
More details about the evaluation of the exchange interactions, in particular in relation to the basis set used for the local orbitals, can be found in Ref.~\onlinecite{Jijs-in-rspt}.

Relativistic effects will not be considered in our work, for sake of simplicity and unless explicitly stated. These effects give rise to other types of magnetic interactions, like anisotropic exchange couplings and magnetocrystalline anisotropy. However, the bilinear term, considered in the present work, is usually the leading one. For instance, it was recently shown for Fe/Rh(001) that by considering Heisenberg interactions only it is possible to obtain a very detailed picture of magnetic excitations, giving an excellent agreement with experiment~\cite{meng}. Nevertheless, we performed a few additional simulations with spin-orbit coupling included in order to analyze the enhancement of the orbital magnetism at the surfaces, which is a very important problem in materials science. These results will be presented at the end of the paper. 

\subsection*{Computational Details}

DFT simulations were performed by using LDA as exchange-correlation functional. 
After the convergence, we have applied a LDA+DMFT technique for a selected set of TM $3d$ orbitals.  
The \textit{k} integration over the irreducible wedge of the Brillouin zone have been performed using 24$\times$24$\times$24 points for bulk and 24$\times$24$\times$1 points for the slabs. 
We have performed relaxation of the topmost layers of Fe slab, which are known to be quite small in these systems~\cite{quinn}. In LDA, for Fe, we obtained a 0.1$\%$ (1$\%$) reduction of the surface (subsurface) magnetic moments with respect to unrelaxed slabs (truncated bulk).  For Co and Ni, we did not perform extensive tests as we expect the changes induced by the relaxation to be even smaller. In fact these changes are proportional to the difference between bulk and surface spin moments, which is much larger for Fe than for Co and Ni.
Therefore, to avoid presenting two sets of similar results and to facilitate comparison with similar studies, our analysis will be limited to unrelaxed slabs, where the interatomic distances depend solely on the bulk lattice parameter. The latter was chosen as obtained from experiments, i.e. 2.86 {\AA} for bcc Fe, 3.52 {\AA} for fcc Ni and 2.51 {\AA} for hcp Co~\cite{marder}. For the latter, the distance between the hexagonal planes was chosen as 4.07 \AA~\cite{marder}.
The free-standing slabs of Fe, Co and Ni have been modeled using 15 layers of their bulk structure repeated in (001) direction for Fe and Ni, while the (0001) direction was used in the case of Co. 
Since three dimensional periodic boundary conditions are used, a 27-\AA-thick layer of vacuum was used to construct a supercell. 

LDA+DMFT simulations were performed for a temperature of 400 K. The effective impurity problem arising in DMFT was solved through the spin-polarized T-matrix fluctuation-exchange (SPTF) solver~\cite{sptf}. Since the latter is a perturbative approach, it can only be applied to systems with moderate correlations and in the metallic regime of the Mott-Hubbard transition. SPTF is usually applied by using the static part of the self-energy as a double-counting correction term. This choice has been used for all DMFT simulations, throughout the paper. 

For $3d$ orbitals, where the electrons are supposed to show more atomic-like features, the Coulomb interaction can be parameterised via Slater integrals $F^n$\cite{stoner}:
\begin{equation}
U = F^0, \qquad J=\frac{F^2+F^4}{14},
\end{equation}
where U is Hubbard parameter and J is Hund's exchange. 
The values of U and J can be either extracted from experiments or calculated from first principles. 
In this work, we have taken their values from the literature~\cite{sptf,u1,u2}.
LDA+DMFT calculations for Fe and Co were done utilizing U = 2.3 eV and J = 0.9 eV, while for Ni U = 3 eV and J = 0.9 eV were chosen. 
In order to see the effect of U on the spectra and exchange parameters, we have performed some test calculations using larger values of U for surface and subsurface atoms, while keeping the previous value unchanged for the inner layers. Although it is known that the choice of U affects the intensity and the position of satellites in the valence band spectra, e.g. as reported in Ref.~\cite{igor}, our results suggest that the exchange parameters are marginally affected by varying the U value. 
This analysis is illustrated in Appendix \ref{u}.

Finally, we performed extensive tests to analyze the role of self-consistence over the electron density in the LDA+DMFT cycle. We focused on spectral functions and exchange interactions for 7-layer slab. These results are presented in Appendix \ref{csc}. Our general conclusion is that updating the electron density in the LDA+DMFT cycle introduces minor corrections for the transition metals surfaces, at least concerning the magnetic properties. In light of these minor changes as well as for computational efficiency, we performed simulations of 15-layer slabs by keeping the electron density unchanged. 

\section{Results}
\label{results}

\begin{figure}[btp]   
 \includegraphics[width=0.38\textwidth]{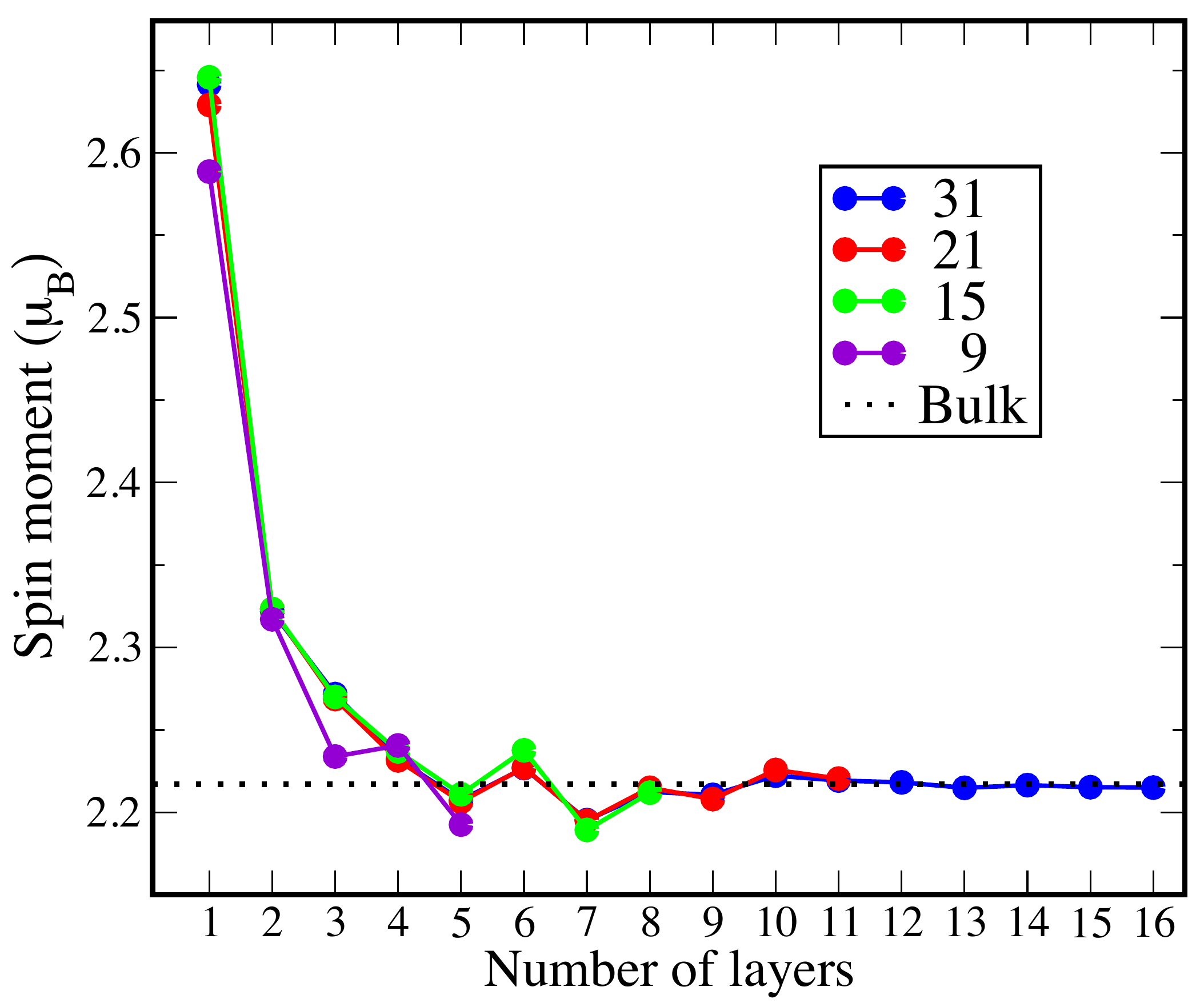}
 \caption{Layer-resolved Fe-$3d$ projected spin moment for different slabs, whose thickness varies from 9 layers to 31 layers. The layers are numbered from 1 (surface) to $N_\text{max}$ (the innermost layer). $N_\text{max}$ is equal to the total number of layers in the slab plus one, divided by two.}
\label{fig:fe_spin}
\end{figure}
We consider free-standing slabs of $3d$ transition metals in their most stable magnetic structure below the Curie temperature. The exchange interaction is calculated for different layers as a function of the interatomic distance between pair of atoms. These calculations are aimed at understanding in detail the differences between bulk and surface, and also at seeing how the local dynamical correlations affect the exchange interactions. In the next few sections we will elaborate on these issues separately for each element.

\subsection{Fe}
\label{fe}
\begin{figure}
\centering
\includegraphics[width=0.47\textwidth]{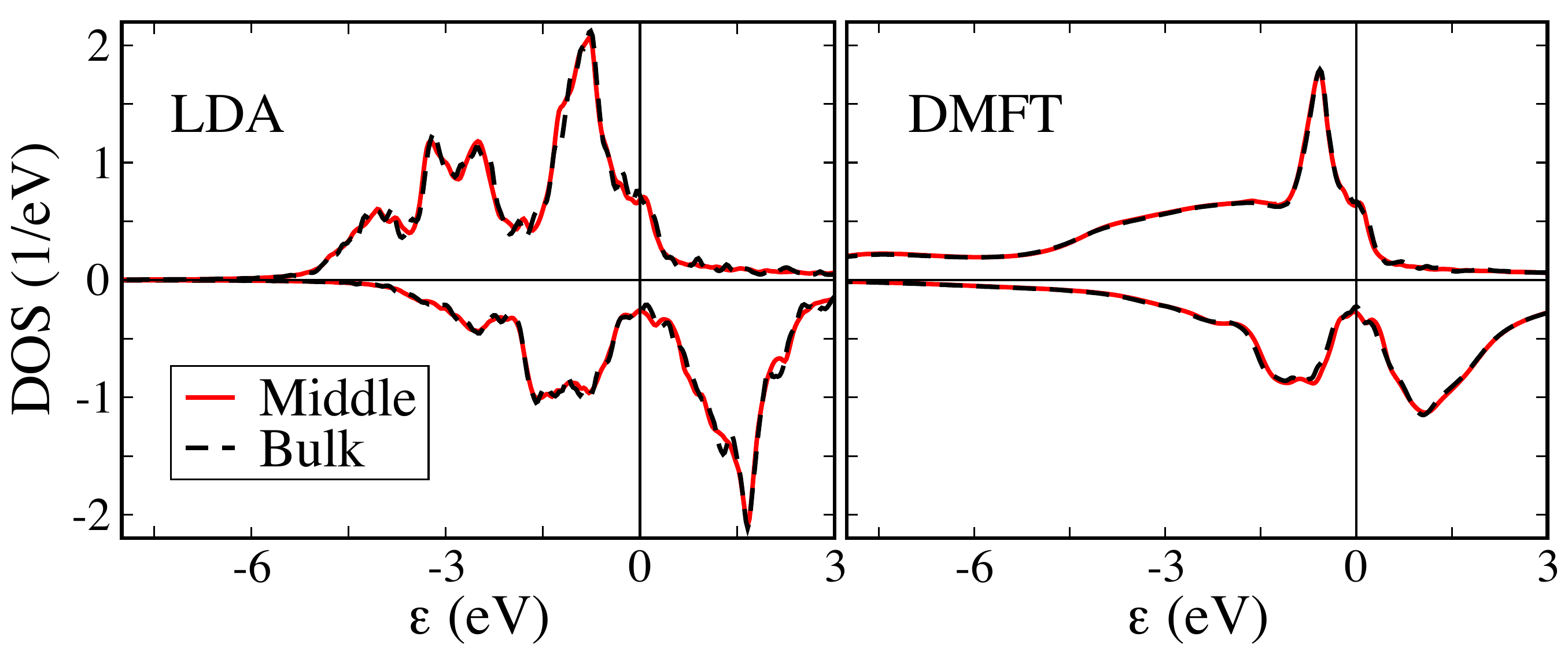}
\caption{Layer-resolved projected density of states of $3d$ orbitals of the innermost layer (middle) of Fe slab and of the bulk for majority and minority spin components in LDA (left panel) and LDA+DMFT (right panel).}
\label{fig:fe_bulk}
\end{figure}

It is important to address first the convergence of the relevant magnetic properties with respect to the thickness of the slab. We present this analysis only for Fe for brevity. In Fig. \ref{fig:fe_spin} layer-resolved Fe-$3d$ projected spin moments are reported for slabs of different thickness. These calculations reveal long range damped oscillations of the local moments when going from the outermost layer to the innermost one. This behavior is due to the surface induced changes in the magnetism of the itinerant ferromagnets (Friedel oscillations)~\cite{solid,friedel}. In principle, the formation of the quantum states in the finite size slabs is accompanied by the creation of a barrier on the surface which leads to a different electronic structure around the Fermi level ($E_F$). In the cases of Fe, Co and Ni, the magnetic moment of the innermost layer reached the bulk value, with a difference smaller than 1$\%$, for slabs of 15 layers. For this size, the exchange interactions of the inner most layer of Co and Ni slabs, were equal to those of their bulk up to $0.5\%$. For Fe, however, a slightly larger difference was obtained, of about $1.2\%$, which was due to the difficulties in matching the same special points for the sampling of two-dimensional and three-dimensional Brillouin zones. Finally, an analogous convergence of the electronic structure can be observed in Fig. \ref{fig:fe_bulk}, where the projected density of states (PDOS) of the innermost layer is compared with the PDOS of the bulk for the 15-layer slab. The curves do not exhibit any visible difference on the scale of interest, for both LDA and LDA+DMFT. 

\begin{figure}[b]
\centering
\includegraphics[width=0.47\textwidth]{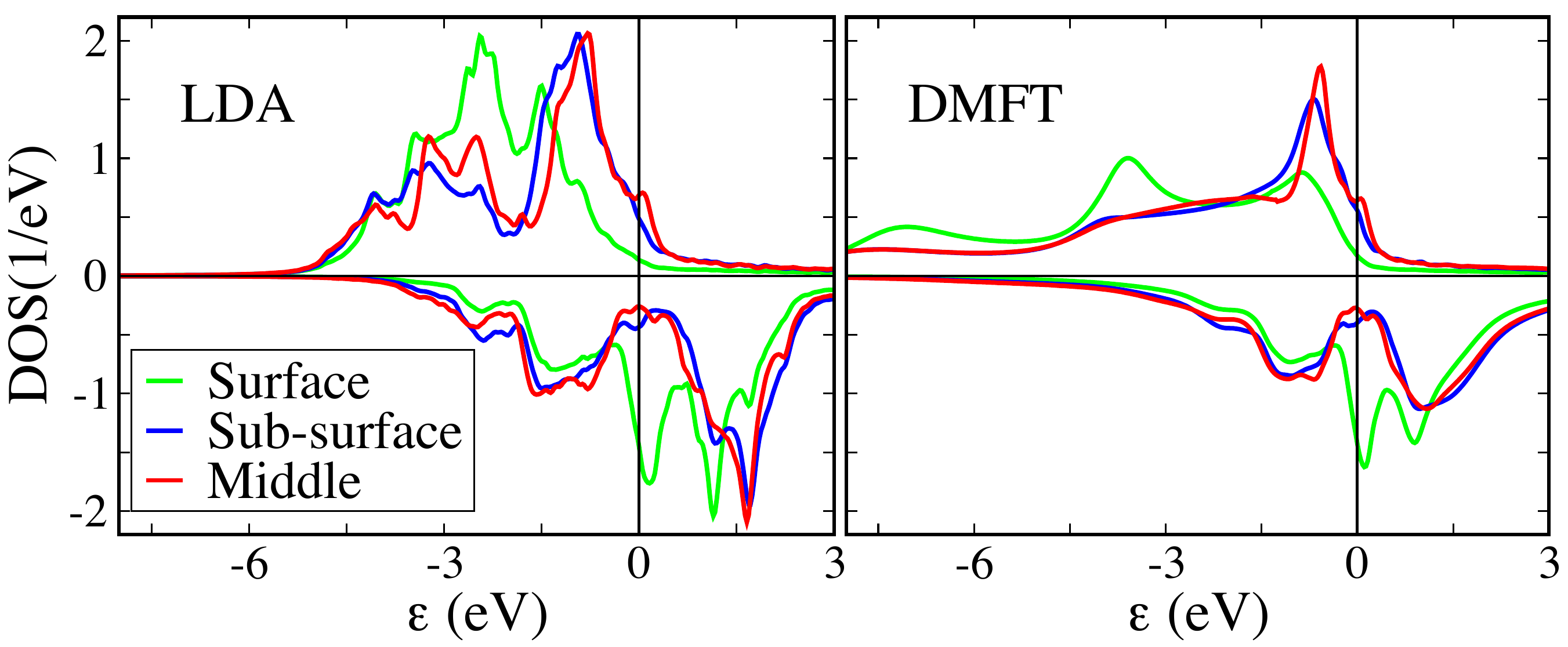}
\caption{Layer-resolved projected density of states of $3d$ orbitals of Fe slab for the atoms at the surface, subsurface and the innermost layer for majority and minority spin components in LDA (left panel) and in DMFT approach (right panel).}
\label{fig:fe_surface}
\end{figure}

Next we have analyzed the differences in the PDOS of the surface atoms and that of the innermost layer. These results obtained in LDA and LDA+DMFT are shown in Fig. \ref{fig:fe_surface}.
The PDOS at the surface is very different from the bulk, due to a reduced coordination number, which results in narrower bands and more pronounced correlation effects. Our results are in a good agreement with prior studies reported by Grechnev \textit{et al.}~\cite{igor} and Chuang \textit{et al.}~\cite{chuang}. Note that in the latter work the calculations were performed using generalized gradient approximation, which accounts for some differences with respect to the results reported here.


Fig.~\ref{fig:fe_surface} shows that there are a large number of majority spin states in the vicinity of $E_F$ for the innermost layer of Fe slab, for both LDA and LDA+DMFT. These states arise mainly due to $d_{yz}, d_{xz}, d_{xy}$ orbitals (not shown here), and are shifted to lower energies for the atoms sitting at the surface. This results into an effective suppression of spectral weight at $E_F$. For minority spin, instead, the bulk (innermost layer) PDOS shows just a few states of the $t_{2g}$ character around $E_F$, where a pseudo-gap forms. At the surface, the increase of the exchange splitting causes these states to move just across $E_F$, which results in a drastic increase of the spectral weight. As a result, the spectral weight at the Fermi level arises mainly from one spin channel, which makes the surface behave as a \textit{strong ferromagnet}. The inner layers have instead the characteristics of a \textit{weak ferromagnet}, similarly to the bulk. From Fig.~\ref{fig:fe_surface}, we also notice that the PDOS for the atoms sitting on the subsurface layer does not show substantial differences with respect to the PDOS of the bulk. 

We can now focus on the comparison between LDA and LDA+DMFT. Although the overall PDOS obtained by means of these two methods are quite different, they exhibit very similar behavior in the vicinity of $E_F$. Given that this region is of primary importance for the exchange interaction, we expect to obtain similar results within these two approaches, at least as concerns the asymptotic behavior.

\begin{figure*}
 \includegraphics[width=0.95\textwidth]{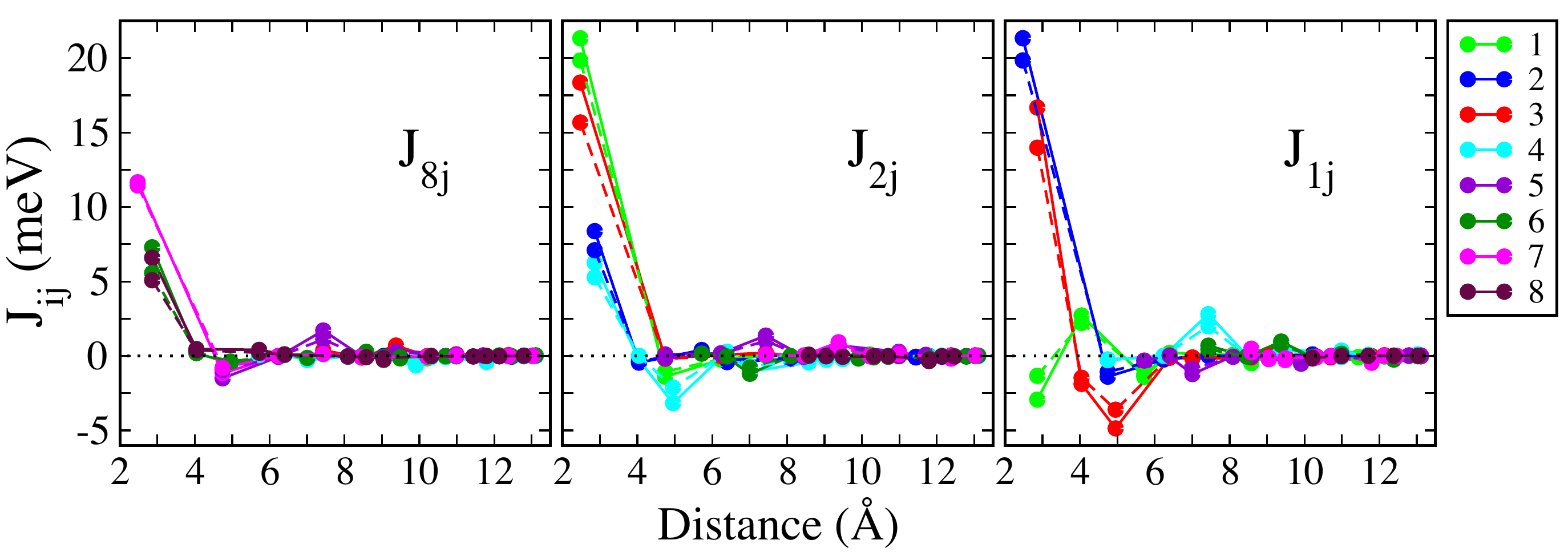}
 \caption{Layer-resolved exchange parameters ($J_{ij}$) for 15-layer bcc Fe(001) slab for the case when atom \textit{i} is located in the innermost layer (left panel), in the subsurface (middle panel) and at the surface (right panel). The solid lines indicate LDA results while the dashed lines represent the LDA+DMFT results. The layer numbering in the legend starts from surface denoted by 1, subsurface denoted by 2 and so on. The innermost layer is denoted by 8. The interaction between the surface and layer 9 and further are small and are not shown in these plots.}
\label{fig:fe_j}
\end{figure*}

From this point on, we focus exclusively on the 15-layer slab. In Fig. \ref{fig:fe_j}, the layer resolved exchange parameters ($J_{ij}$) are reported, for both LDA and LDA+DMFT. For clarity, we report only results for the most physically interesting layers, e.g. surface, subsurface and the innermost layer. We will anyway make general considerations regarding all layers in the following discussion. The \textit{intralayer} exchange interaction is referred to the case when the two atoms interacting with each other are located in the same layer. The \textit{interlayer} interaction is referred to the case when the two atoms belong to different layers. The layers in the plots are denoted by 1 for the surface, 2 for the subsurface and so on, analogously to Fig.~\ref{fig:fe_spin}.

The first general consideration to draw from our calculations is that interlayer exchange parameters for atoms in the inner layers are substantially smaller than those for atoms in layers closer to the surface. This trend is observed for both LDA and in LDA+DMFT. For instance, Fig. \ref{fig:fe_j} shows that the exchange interaction between an atom at the surface and its first nearest neighbor (NN) sitting in the subsurface (blue lines in the right panel) is strongly ferromagnetic. The strength of this exchange interaction is about twice larger than that of an atom in the innermost layer and its first NN in an adjacent layer (pink lines in the left panel of Fig. \ref{fig:fe_j}). 

The quantitative explanation for the layer dependence of the exchange interactions can be provided by the two important factors. First, there is a direct influence of the on-site exchange field at the atoms $i$ and $j$ on the corresponding $J_{ij}$ parameter (see Eq.~(2)).
The second important factor is the coordination numbers which would affect the Hamiltonian $\hat{H}$, the self-energy and, therefore, the inter-site Green's function (see Eq.~(3)). 

An inspection of Fig.~\ref{fig:fe_surface} has already revealed that majority and minority spin states are more split for the surface atoms. Hence, the local exchange field ($\Sigma^s$) is overall larger than its bulk counterpart, which can explains the enhancement of the exchange integrals between surface and subsurface atoms with respect to couplings between more internal (adjacent) layers. Moreover, the surface PDOS is characterized by a larger spin polarization of the Fermi surface in comparison to inner layers. This provides a large number of available states with a certain spin projection right above $E_F$. Hence, similarly to the double exchange mechanism, an electron hopping will be facilitated if it does not have to flip its spin, i.e. if the neighboring moments are parallel to each other. This scenario thus also supports an enhancement of the ferromagnetic interaction with the surface atoms, as obtained in our calculations.

We then proceed to the analysis of the intralayer coupling. An interesting finding of the present investigation is that two Fe atoms at the surface possess an AFM coupling (green lines in the right panel of Fig \ref{fig:fe_j}). Note that this neighborhood corresponds to second NN atoms in bcc structure, and is therefore always smaller than the leading FM contribution between first NN. The presence of AFM exchange interactions at the surface is rather surprising, albeit we note that previous first principles theory also suggest such a coupling~\cite{turek}. Fe is known to have the tendency to AFM coupling for hcp or fcc crystal structures~\cite{FeW-STM,TM-on-W-Blugel,freeman,kudrnov} as well as for thin mono-layers on some substrates, but seldom in bcc like environments~\cite{chuang, meng, deak}. For example, the exchange coupling between the two neighboring atoms at the surface of Fe clusters was reported to be FM~\cite{minar}.

Here we focus on the understanding the origin of this tendency to AFM coupling at the surface. 
For this purpose we analyze individual orbital contributions to the exchange parameter. 
The local Hamiltonian for each Fe atom is diagonal in the basis of cubic harmonics and so is $\Sigma^s$ from Eq.~\ref{eq:jij}. Having $\Sigma^s$ in a diagonal form allows us to write each exchange coupling as $J_{12}=\sum_{m_1,m_2} J^{m_1,m_2}_{12}$, where orbital $m_1$ is located at the site 1 and orbital $m_2$ is at the site 2. Exchange interaction between two closest surface spins in a form of a matrix in orbitals space is shown in Table \ref{tab:table2}. The table hence shows the strength of the exchange interactions of symmetry resolved states of one atom with symmetry resolved states of a nearest neighbor surface atom. 
The total interaction between these two atoms is obtained by summing all components of Table I. 
The analysis reveals that there are basically two competing contributions to the $J_{ij}$ between the NN's surface moments.
A first FM contribution originates from $d_{xy}-d_{xy}$ bonds and $d_{yz}-d_{yz}$ bonds, depending on the bond vector. 
A second AFM contribution, instead, arises from $d_{yz}-d_{x^2-y^2}$ bonds. 
This contribution is much stronger in our case, and overcomes the FM part of the exchange. 
Similar competition was shown to take place for the next NN exchange couplings in bulk Fe, which corresponds to the same coordination shell as we investigate here.\cite{Jijs-in-rspt}
However, the balance between the two contributions can be easily changed by small changes of the NN distance. 
For example, Chuang \textit{et al.}~\cite{chuang} reported a FM coupling between the adjacent surface atoms in free standing Fe films when using the Ir lattice constant. We have repeated their calculations and obtained the same results. We conclude that the tendency to AFM coupling is innate in the magnetic properties of the Fe surface. This suggests that the primary role of the substrate consists in modifying inter-atomic distances and not in affecting the electronic structure via direct hybridization, at least for the aforementioned cases.

\begin{figure}
\centering
\includegraphics[width=0.47\textwidth]{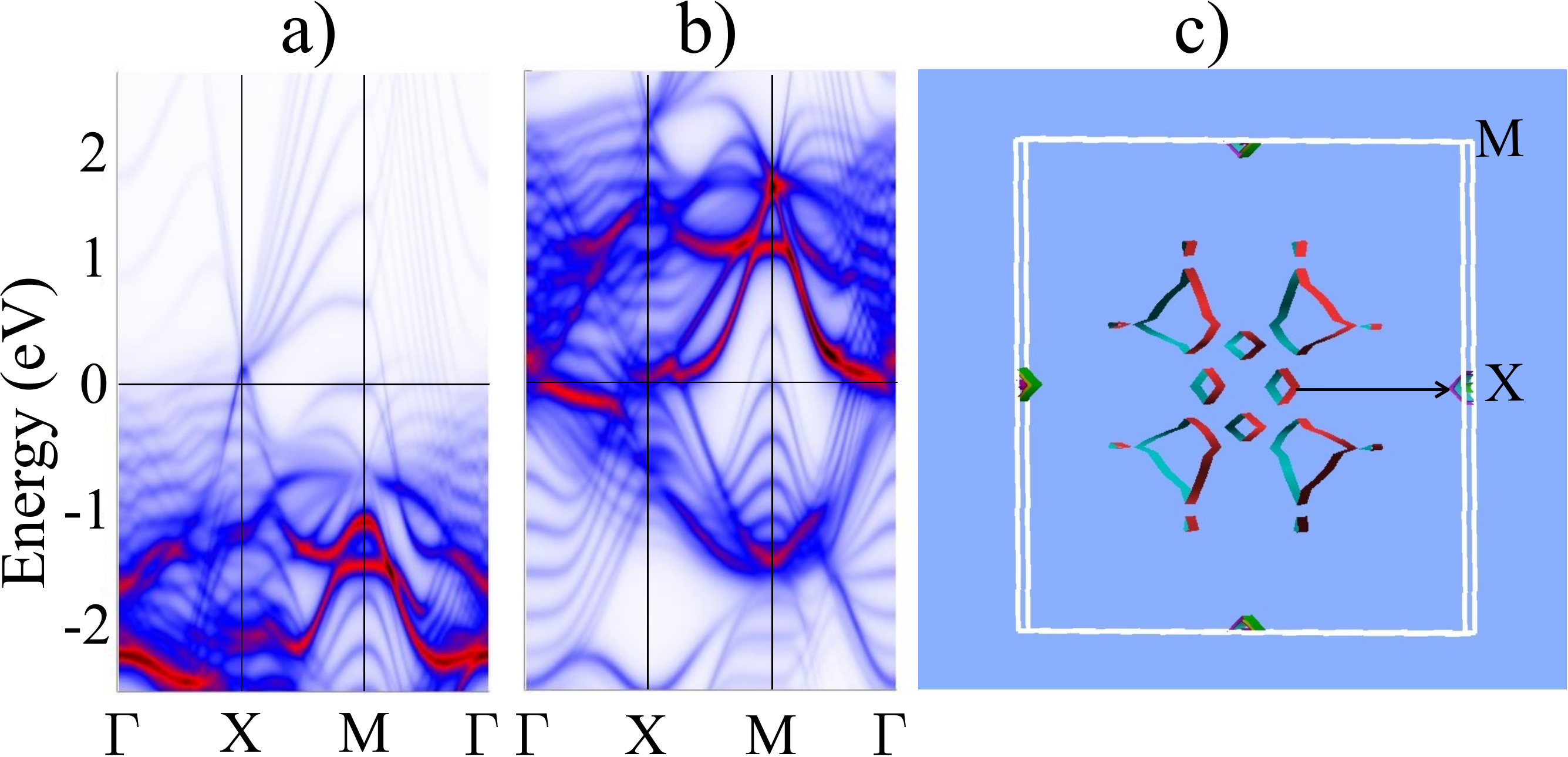}
\caption{Panels (a) and (b): the calculated band structure of Fe slab for majority and minority-spin states, respectively, shown with the amount of $3d$ orbital character of the surface atoms (red color). Panel (c): The Fermi surface cross-section of the minority-spin bands, having the largest contributions from surface $3d$ states. Suggested nesting vector is indicated with the black arrow (the spectra are calculated in the complex energy $E+i\delta$ with $\delta$=0.005 Ry).}
\label{fig:fe-fermi}
\end{figure}

In order to have a more clearer explanation of the AFM interactions at the Fe surface, we have studied its underlying electronic structure in more detail.
In Fig.~\ref{fig:fe-fermi}(a,b) we show the spin-polarized band structure, projected onto $3d$ states of the surface atoms. 
The largest contribution of these orbitals to each band is shown with the red colour. 
As one can see from Fig. 5, the surface states contribute only to the majority-spin bands well below $E_F$. 
In contrast, there is a significant weight of them at $E_F$ for the minority-spin channel, in particular near $\Gamma$ and $X$ points, which will contribute to the magnetic susceptibility.
To strengthen this point, in Fig.~\ref{fig:fe-fermi}(c) we show parts of the Fermi surface cross-section, originating from the bands with a large surface component.
One can see that the Fermi surface is nested between $\Gamma$ and $X$ points and the nesting vector is indicated with an arrow in Fig.5(c). 
The nesting vector (indicated with the black arrow) is directed along (100) direction and has a length very close to $\pi/a$.
This defines a preferable direction for a symmetry-breaking in the system and we suggest it is connected to the AFM coupling between the NN spins at the surface.
We note as well that bulk bcc Fe has similar features of the Fermi surface (see e.g. Fig.~3 in Ref.~\onlinecite{fermi}).
However, the nesting vector connecting the pockets located at $\Gamma$ and $H$ points has much a smaller length, which does not lead to the pronounced AFM interaction.

\begin{table}[bp]
\caption{\label{tab:table2} Orbital-decomposed exchange interaction parameter between two closest neighbors at the surface for the bond vector (010). The reported values are in meV.}
\begin{ruledtabular}
\begin{tabular}{lrrrrr}
 & \textrm{$d_{z^2}$\:} & \textrm{$d_{x^2-y^2}$}\hspace{-0.2cm} & \textrm{$d_{yz}$\:} & \textrm{$d_{xz}$\:} & \textrm{$d_{xy}$\:} \\
\hline
$d_{z^2}$& 0.81 & -1.02 & 0.36
& 0.00 & 0.00 \\
$d_{x^2-y^2}$& -1.02 & 1.26 & -6.26
& 0.00 & 0.00 \\
$d_{yz}$& 0.36 & -6.26 & 3.99
& 0.00 & 0.00 \\
$d_{xz}$& 0.00 & 0.00 & 0.00
& -1.07 & 0.05 \\
$d_{xy}$& 0.00 & 0.00 & 0.00
& 0.05 & 6.32\\
\end{tabular}
\end{ruledtabular}
\end{table}

Other considerations can be drawn from the calculated exchange parameters from Fig. \ref{fig:fe_j}. AFM intralayer coupling between the closest neighboring atoms at the surface (light-green lines in the right panel) might in principle lead to non-collinear spin configurations. However, we estimate the resulting frustration to be weak, since their interlayer couplings with atoms in the substrate are much larger (dark-blue lines in the right panel).

At the surface, the AFM exchange coupling between the closest neighbors as well as the lower coordination number result in smaller values for the total exchange interactions $J_1=\sum_j J_{1j}$. This consequence is, qualitatively, in agreement with the LMTO-ASA results reported by Turek \textit{et. al}~\cite{turek}. A quantitative comparison is, however, not possible for Fe, for which the long-range oscillatory (RKKY-like) behavior of exchange constants makes the value of $J_1$ strongly dependent on the number of shells taken into account.

Overall, both LDA and LDA+DMFT deliver very consistent results for the exchange parameters, although the latter are slightly smaller. Once the dynamical correlations are introduced, as long as the topology of the Fermi surface is unchanged, the main effect is to produce carrier mass renormalization. 
A qualitative explanation for the overall decrease of the $J_{ij}$'s in LDA+DMFT can be found in Ref.~\onlinecite{mvv-j-rescale}, where a direct link between the total exchange coupling $J_{i}$ and the renormalisation factor $Z$ is established.
However, in a multi-orbital case, different (by symmetry) orbitals are characterized by different renormalization factors $Z$ and therefore there is no simple scaling relation between the overall exchange couplings extracted from LDA and LDA+DMFT. In Appendix \ref{qw} we show the computed $Z$-factors for each $3d$ orbital centered on the atom sitting either in the surface, subsurface or in the middle layer. One can see that in case of Fe slab the $Z$-factors for the majority and minority states are very different, reaching the maximal difference for surface electrons (being 0.53 and 0.73, respectively).
A more detailed information about the mass enhancement as a measure of the strength of the many-body effects in all studied surfaces can be found in Appendix \ref{qw}.

\subsubsection{Different surface directions of bcc Fe}
\begin{figure}
\centering
\includegraphics[width=0.45\textwidth]{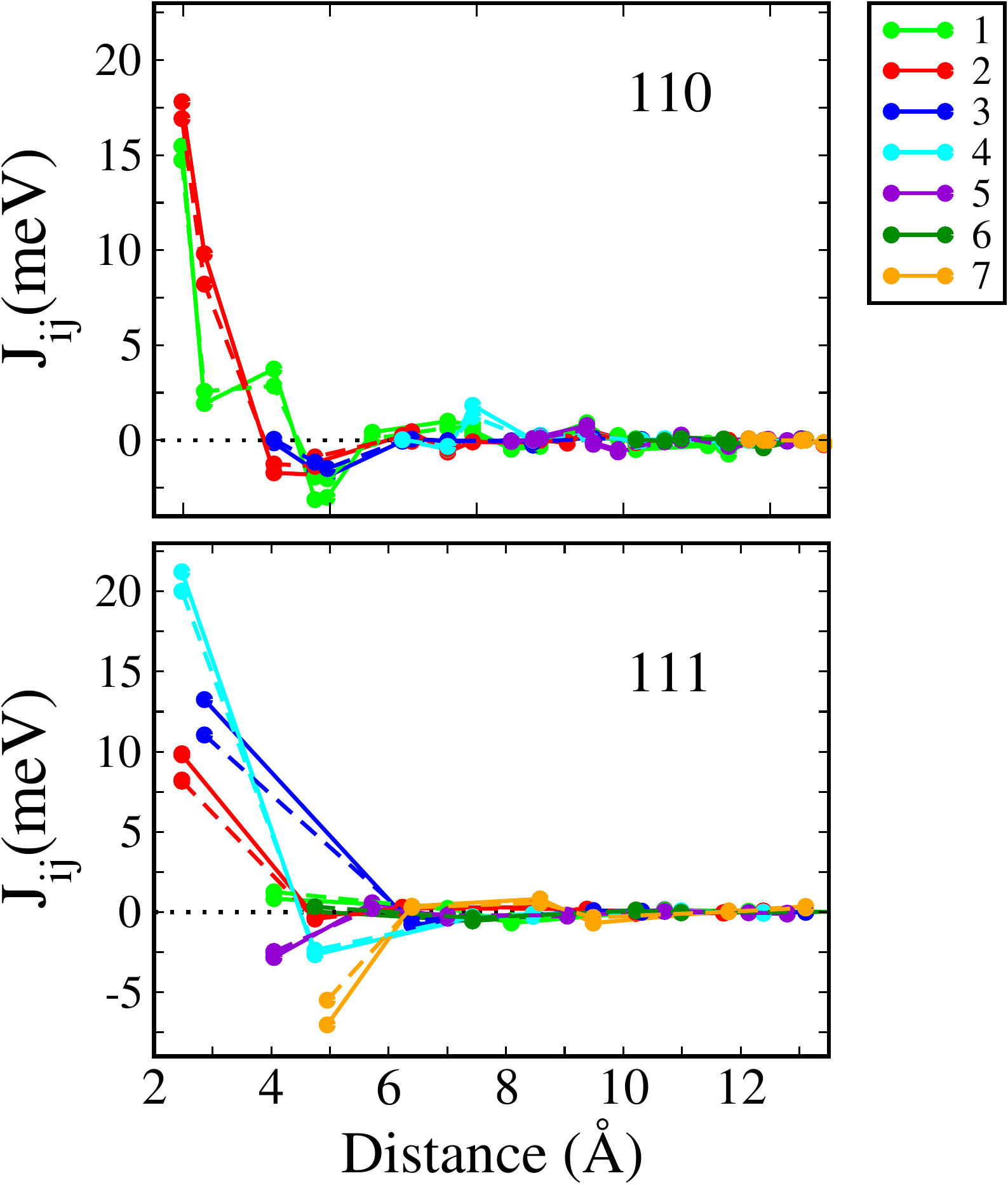}
\caption{Layer-resolved exchange parameters for bcc Fe slab in (110) and (111) directions (top and bottom panels respectively). Here, we only show the results for the case when atom \textit{i} is located at the surface ($J_{1j}$) . The solid lines indicate LDA results while the dashed lines represent LDA+DMFT results. The layer numbering in the legend starts from surface denoted by 1, subsurface denoted by 2 and so on.}
\label{110-111}
\end{figure}

In order to see whether the AFM exchange coupling between the two nearest neighbors at the surface of Fe can be observed in different surface directions, we have performed some additional calculations for the directions of (110) and (111) of bcc Fe. The obtained exchange parameters between the atom at the surface and the ones in any layer ($J_{1j}$) are shown in Fig.~\ref{110-111}. As clear from the left panel of Fig.~\ref{110-111}, the strong FM coupling happens for the NN atoms at the surface of (110) direction as well as between the second and third NNs. In contrast, such a strong coupling has not been observed between the nearest atoms at the surface of (111) direction. The reason is that in this case, the closest atoms at the surface are the fourth NN of each other, too far to show significant coupling.
As a conclusion, our calculations show that the AFM coupling can only be seen at the surface of (001) direction.  

\begin{figure}[bp]
\centering
\includegraphics[width=0.47\textwidth]{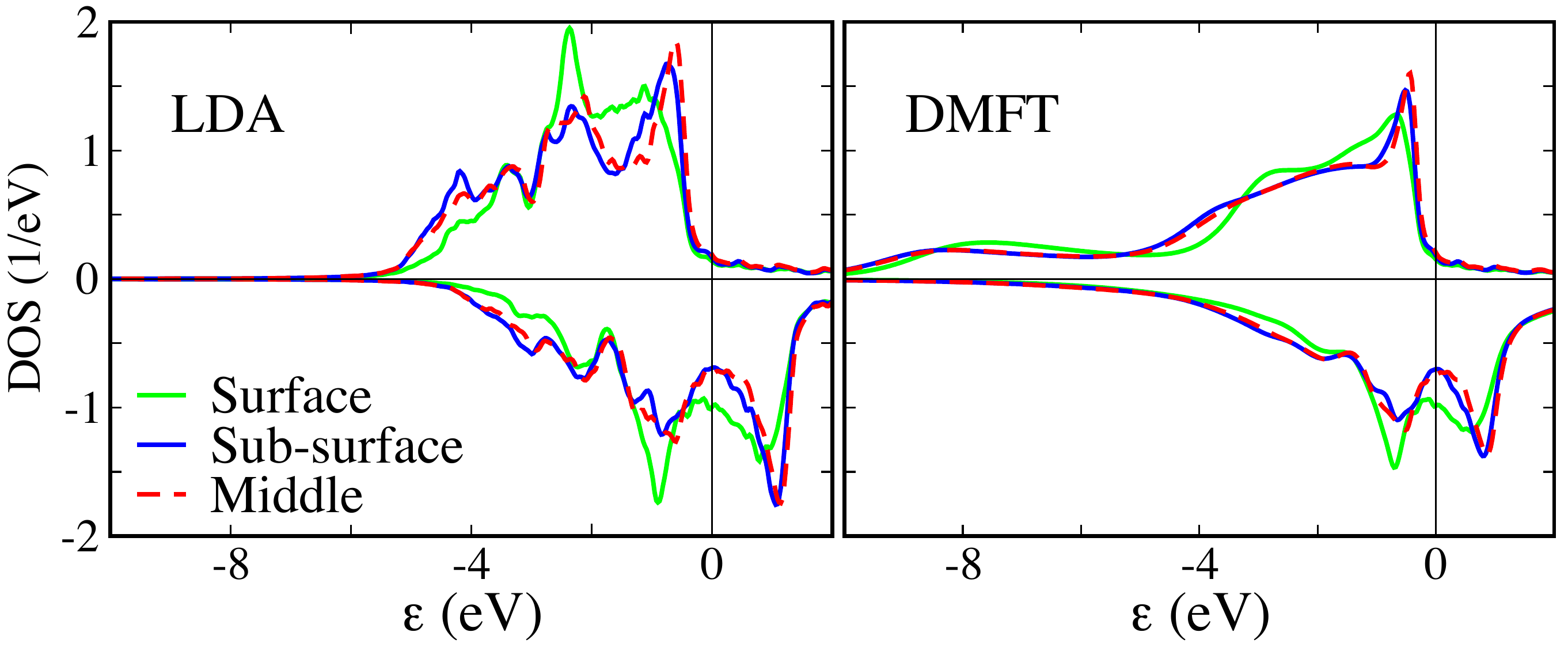}
\caption{Layer-resolved projected density of states of $3d$ orbitals of Co slab for the atoms sitting at the surface, subsurface and the innermost layer for majority and minority spin components in LDA (left panel) and in DMFT approach (right panel).}
\label{fig:co_dos}
\end{figure}

\begin{figure*}
 \includegraphics[width=0.95\textwidth]{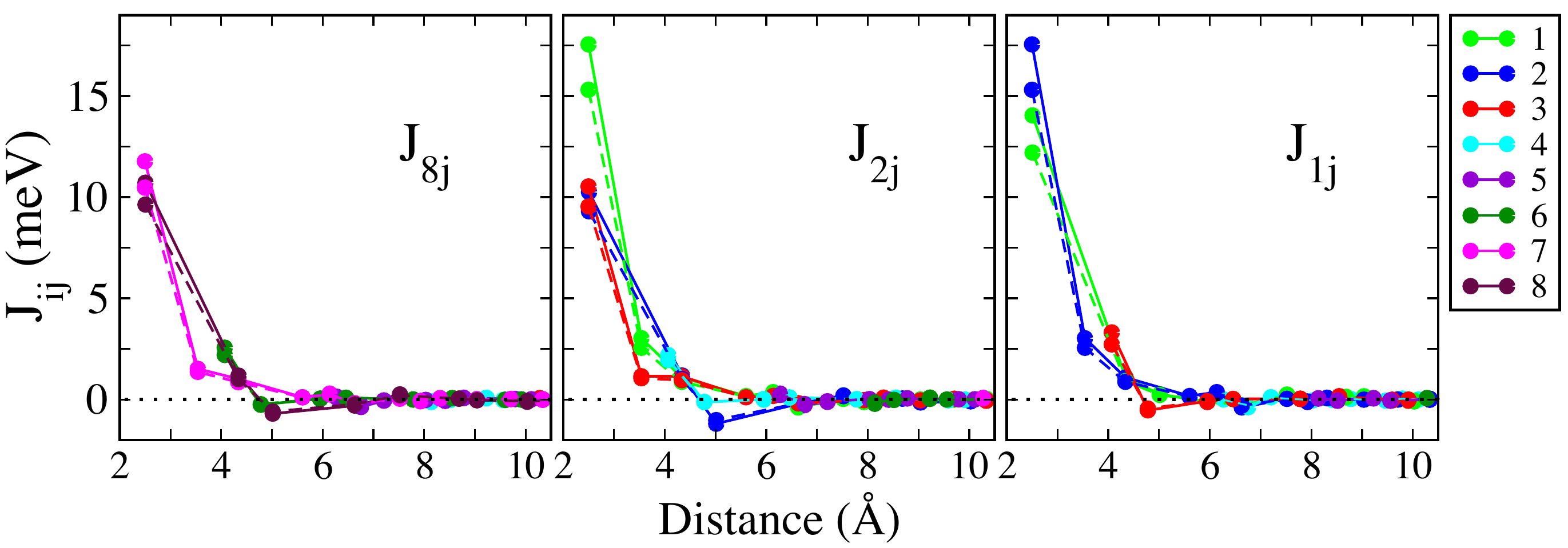}
 \caption{Layer-resolved exchange parameters ($J_{ij}$) for 15-layer hcp Co (0001) slab for the case when atom \textit{i} is located in the innermost layer (left panel), in the subsurface (middle panel) and at the surface (right panel). The solid lines indicate LDA results while the dashed lines represent the LDA+DMFT results. The layer numbering in the legend starts from surface denoted by 1, subsurface denoted by 2 and so on. The innermost layer is denoted by 8. The interaction between the surface and layer 9 and further are small and are not shown in these plots.}
\label{fig:co_j}
\end{figure*}

\subsection{Co}
\label{co}

Next, we have considered a slab of hcp Co containing 15 layers repeated in the (0001) direction.  
In Fig. \ref{fig:co_dos} the PDOS for the innermost layer, subsurface and surface atoms are shown. The results obtained through both LDA and LDA+DMFT are reported.  
One can see that the PDOS for atoms in the innermost layer and in the subsurface are similar around the Fermi level, although their overall shapes are slightly different. 
For instance, the first peak below $E_F$ for majority spin states has lower intensity and located at lower energies for atoms at the subsurface, in comparison with atoms in the innermost layer. 


Similar conclusions can be drawn for the surface. The majority spin PDOS for atoms at the surface shows quite similar features around $E_F$ as for atoms in the innermost layer. This means that the strong ferromagnetic behavior is found both for the surface of Co like as well as the bulk. Nevertheless, the sharp peak below the Fermi level, which mostly arises from $d_{x^2-y^2}$ and $d_{xy}$ contributions, is slightly suppressed and shifted to lower energies for atoms at the surface. Conversely, at the surface, there is a slight increase in the spectral weight around the Fermi level for the minority spin states. This weight is mainly due to orbitals with $d_{yz}$ and $d_{xz}$ symmetry. 
In addition, as we saw for the case of Fe, this will result in larger values for the local exchange field ($\Sigma^s$) at the surface and consequently in larger exchange integrals. Finally, Fig. \ref{fig:co_dos} also shows that the PDOS obtained from LDA and LDA+DMFT show similar features around the Fermi level. Hence, one would expect a similar trend in the asymptotic behavior of the exchange parameters obtained within these two methods. 

In Fig. \ref{fig:co_j}, layer-resolved exchange parameters are displayed for the physically most interesting layers. Results from both LDA and LDA+DMFT are reported. In contrast to the Fe slab, a relatively faster decay of the exchange parameters can be seen for the Co slab. This is due to that the RKKY character is less effective in strong magnets, e.g. as pointed out in Ref.~\onlinecite{pajda}. However, itinerant magnets, in general, are not perfect strong magnets, due to the hybridization between \textit{d} orbitals and \textit{sp} states. Our statement of strong ferromagnetism in this paper should be viewed as describing a situation when the majority spin states of the DOS is completely filled and hence pushed below the Fermi level.

In agreement with the results for the Fe slab, we obtained that the \textit{interlayer} exchange parameters between NNs are substantially smaller in the inner layers than in layers close to the surface (for comparison, see the green lines in the middle panel of  Fig. \ref{fig:co_j} and the pink lines in the left panel). However, in contrast to the Fe surface, there is a strong \textit{intralayer} FM coupling between the NN at the Co surface, both in LDA and in LDA+DMFT (light green lines in the right panel). An analysis of individual orbital contributions to these exchange parameters reveals the there are strong FM contributions arising from all $3d$ orbitals where the $d_{yz}-d_{yz}$ and $d_{xy}-d_{xy}$ contributions are the strongest.

Despite the strong intralayer and interlayer FM coupling between atoms at the surface, the associated total exchange interaction ($J_1$) is still smaller than those obtained for the inner layers, due to the lower coordination number. As for Fe, the consequence is, qualitatively, in agreement with the conclusions reported by Turek \textit{et. al}~\cite{turek}, but the magnitude is very dependent to the number of shells included in the calculation of $J_1$.

As seen in Fig. \ref{fig:co_j}, both LDA and LDA+DMFT approaches deliver quite similar results for the exchange parameters. The only difference, as we saw for the case of Fe,  is the reduction in magnitude of the $J_{ij}$'s obtained within LDA+DMFT approach, while the overall behavior is similar and the sign is the same. 

\subsection{Ni}
\label{Ni}

\begin{figure}[bp]
\centering
\includegraphics[width=0.47\textwidth]{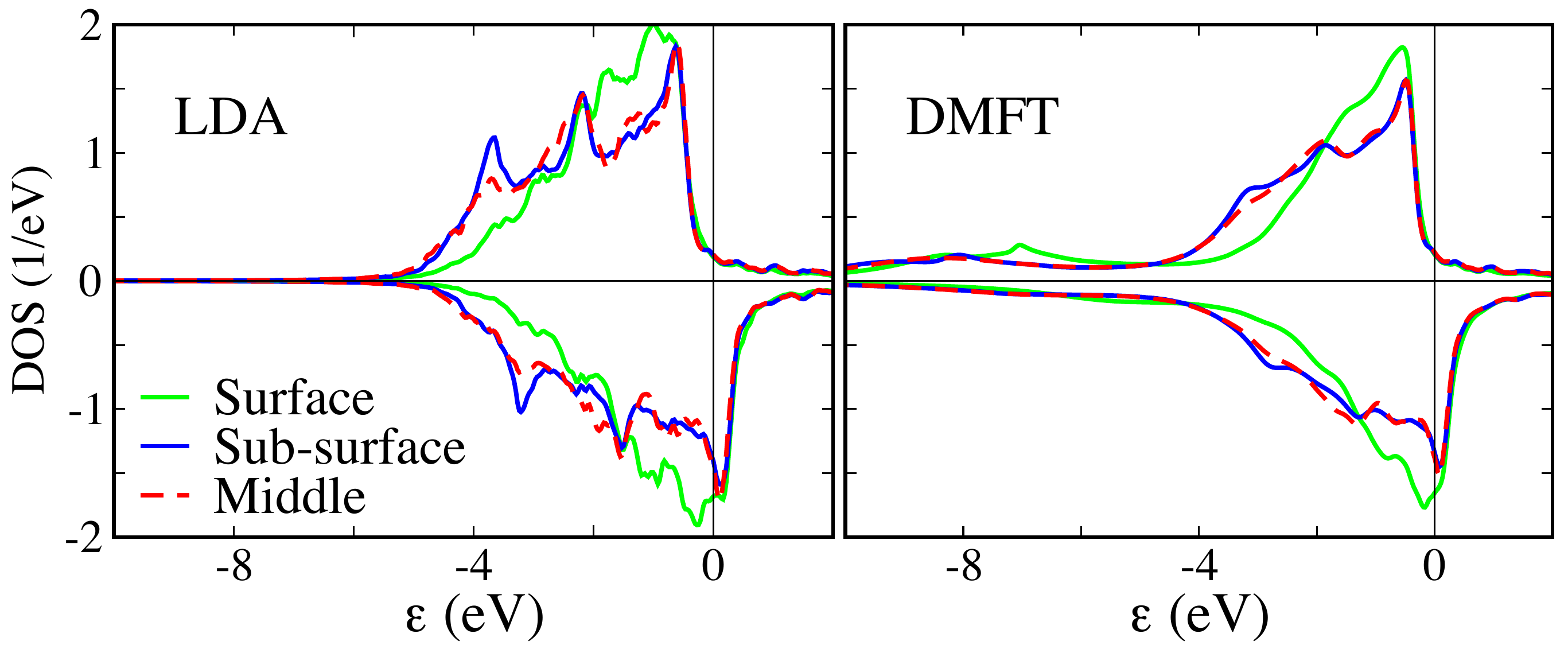}
\caption{Layer-resolved projected density of states of $3d$ orbitals of Ni slab for the atoms sitting at the surface, subsurface and the innermost layer for majority and minority spin components in LDA (left panel) and in DMFT approach (right panel).}
\label{fig:ni_dos}
\end{figure}

\begin{figure*}
\includegraphics[width=0.95\textwidth]{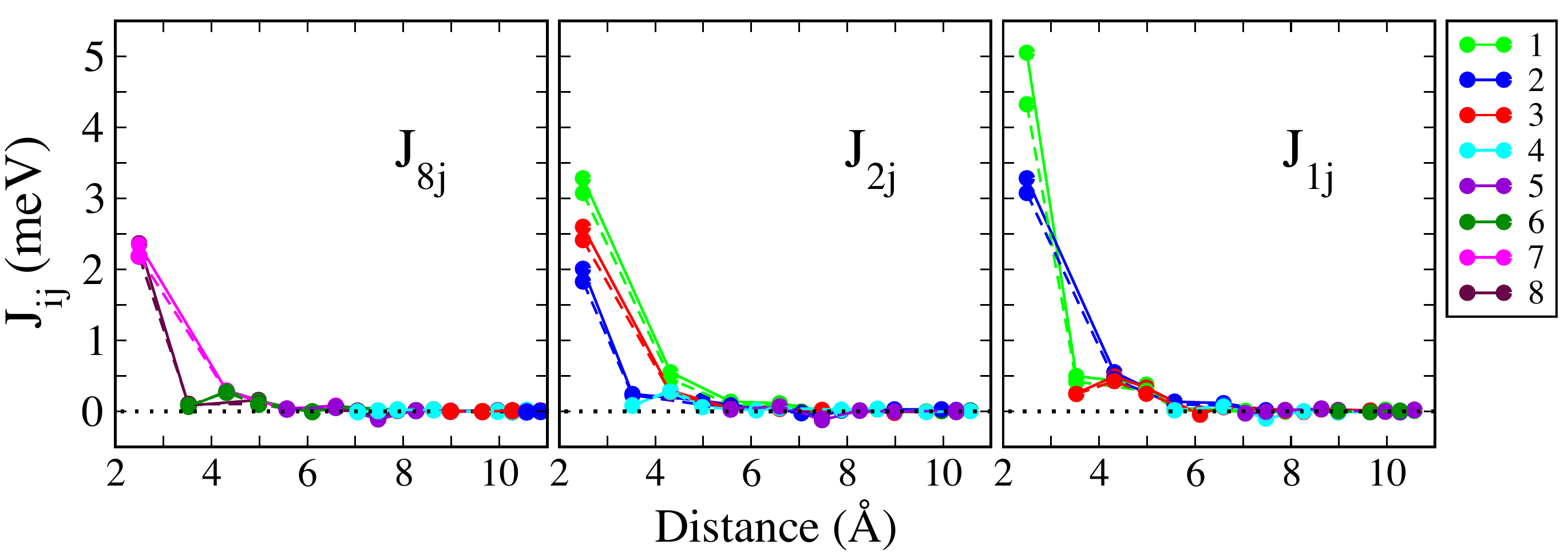}
 \caption{Layer-resolved exchange parameters ($J_{ij}$) for 15-layer fcc Ni(001) slab for the case when atom \textit{i} is located in the innermost layer (left panel), in the subsurface (middle panel) and at the surface (right panel). The solid lines indicate LDA results while the dashed lines represent the LDA+DMFT results. The layer numbering in the legend starts from surface denoted by 1, subsurface denoted by 2 and so on. The innermost layer is denoted by 8. The interaction between the surface and layer 9 and further are small and are not shown in these plots.}
\label{fig:ni_j}
\end{figure*}

Finally, we have considered a slab of fcc Ni consisting of 15 layers repeated in the (001) direction. In Fig. \ref{fig:ni_dos} the PDOS for atoms in the innermost (middle), subsurface and surface layers are reported, for both LDA and LDA+DMFT. The PDOS for an atom in the innermost layer is similar to that of an atom at the subsurface, specially at the vicinity of $E_F$. Discrepancies are visible at higher binding energies, where the peaks become narrower at the subsurface. These differences are small. Around the Fermi level, the PDOS of the majority spin for atoms at the surface is similar to that for the innermost layer. 

Minority spin states, instead, show some differences between atoms at the surface and in the innermost layer. These differences originate mainly from that the surface exhibits a larger contribution of $d_{z^2}$ and $d_{x^2-y^2}$ states to the $E_F$. 

Similarly to Fe and Co, the PDOS obtained via LDA and LDA+DMFT are rather similar around the Fermi level, but possess stronger differences at higher excitations energies. 

Layer-resolved exchange parameters for Ni are reported in the top panel of Fig. \ref{fig:ni_j}, for both LDA and LDA+DMFT. Similar to the case of hcp Co, a relatively fast decay in exchange parameters with distance has been observed. This is consistent with the less pronounced RKKY character reported for bulk strong ferromagnets~\cite{pajda}. However, the magnitude of the coupling for Ni is obtained to be about three times smaller than those for Co. Like in the case of Fe and Co, the NN interlayer exchange parameters are larger for the layers close to the surface (for comparison see the pink lines in the left panel and green lines in the middle panel). As we have seen for the Co surface, there is a FM in-plane exchange coupling between the NN at the surface of Ni both in LDA and LDA+DMFT approaches (green lines in the right panel).

\begin{table*}
\caption{\label{tab:table} Layer-resolved spin ($\mu_s$) and orbital moment ($\mu_o$) for Fe, Co and Ni slabs using LDA (DMFT) approach including spin-orbit coupling corrections as well as the experimental values.}
\begin{ruledtabular}
\begin{tabular}{lcccccc}
 &\multicolumn{2}{c}{Fe}&\multicolumn{2}{c}{Co}&\multicolumn{2}{c}{Ni}\\
 &$\mu_s^{calc}$& $\mu_l^{calc}$ & $\mu_s^{calc}$ & $\mu_l^{calc}$ &  $\mu_s^{calc}$ & $\mu_l^{calc}$ \\ 
 \hline \\ [-1.5ex]
\textrm{Surface}&2.92 (2.94)&0.107 (0.122)
&1.75 (1.79)&0.086 (0.122)
&0.75 (0.77)&0.066 (0.080)\\
 Subsurface&2.32 (2.36)&0.057 (0.067)
 &1.68 (1.73)&0.077 (0.111)
 &0.67 (0.68)&0.055 (0.067)\\
 Middle&2.20 (2.24)&0.051 (0.063)
 &1.65 (1.70)&0.076 (0.108)
 &0.63 (0.65)&0.047 (0.057)\\ [1ex]
 \hline \\[-1.5ex] 
 &$\mu_s^{calc}$& $\mu_l^{calc}$ & $\mu_s^{calc}$ & $\mu_l^{calc}$ &  $\mu_s^{calc}$ & $\mu_l^{calc}$ \\ 
Bulk &2.15\footnotemark[1]&0.080\footnotemark[1]&1.52\footnotemark[1]&0.140\footnotemark[1] &0.51\footnotemark[1] &0.043\footnotemark[1] \\
Bulk &2.08\footnotemark[2]&0.092\footnotemark[2] &1.52\footnotemark[2] &0.147\footnotemark[2] &0.52\footnotemark[2] &0.051\footnotemark[2]  \\
Bulk &1.98\footnotemark[3]&0.085\footnotemark[3] &1.62\footnotemark[3] &0.154\footnotemark[3]&0.65\footnotemark[4]&0.055\footnotemark[4]  \\
Bulk &&&1.86\footnotemark[5]&0.130\footnotemark[5]\\
Bulk &&&1.72\footnotemark[6]&0.134\footnotemark[6]\\
Surface &&&1.92\footnotemark[6]&0.234\footnotemark[6] \\
\end{tabular}
\begin{tablenotes}
\item[a] Ref~\cite{stearn}. \quad
\item[b] Ref~\cite{reck}. \quad
\item[c] Ref~\cite{chen}. \quad
\item[d] Ref~\cite{moon2}. \quad
\item[e] Ref~\cite{moon1}. \quad
\item[f] Ref~\cite{tischer}. These values are for fcc Co. \quad \quad \quad \quad \quad
\end{tablenotes} 
\end{ruledtabular}
\end{table*}

Interestingly, the sum of all exchange parameters for atoms at the surface ($J_1$) is significantly larger than the corresponding sum ($J_2$) at the subsurface, for both LDA and LDA+DMFT. This amount is 47.96 (41.87) meV in LDA (LDA+DMFT) for the surface versus 46.36 (41.35) meV for the subsurface. This might seem to be in contrast to the fact that a lower coordination number should lead to a lower total exchange parameter. However, equation \eqref{eq:jij} shows that the dependence on the exchange splitting is more relevant, which explains our results for Fe, Co and Ni. We should also mention that here the total exchange parameters are evaluated inside a shell of 10 {\AA} radius. 

\subsection*{Spin and orbital moments}
\label{soc}

As mentioned above, all calculations presented so far were performed in the scalar relativistic limit. It is interesting to analyze the influence of spin-orbit coupling on the magnetic properties of Fe, Co and Ni surfaces. Therefore, we have performed additional relativistic calculations including spin-orbit coupling corrections, whose results are reported in Table \ref{tab:table}. Our results for the innermost layers are in qualitative agreement with a prior study for bulk~\cite{orbital1} and for the surface~\cite{eriksson1,alden, ain}. Comparing the results of the Fe slab in Table \ref{tab:table} between LDA and LDA+DMFT approaches, reveals that in general DMFT tends to improve the results of LDA for the slabs both in the bulk-like and in the surface regions. This improvement is not only on the magnitude of spin and orbital moments but also in their ratio ($\mu_l/\mu_s$) which for bcc Fe is about 0.023 (0.028) for LDA (DMFT), as can be deduced from Table \ref{tab:table}. Experimental value reported for bcc Fe bulk is about 0.037~\cite{stearn}. For the surface spins, the enhancement of $\mu_l/\mu_s$ is even larger (0.033 in LDA and  0.046 in DMFT) thanks to the more pronounced orbital polarizations at the surface rising from more localized states. For hcp Co and fcc Ni similar conclusions can be drawn. DMFT enhances the value of the orbital moment, both for bulk and surface atoms. For fcc Ni the experimental values of the bulk are in good agreement with theory, while for Fe and Co theory underestimates the value of the bulk orbital moment with 15-25$\%$. However,  DMFT provides a systematically better approach to investigate the orbital moments of these materials, at least judging from the bulk values. Unfortunately surface orbital moments are not frequently reported for these materials, and we list in Table II one measured value of fcc Co (on a Cu 001 substrate) that shows enhancement compared to bulk values. On the other hand, spin moments obtained from LDA are marginally modified by the dynamical correlations.

Finally, we should mention that our results for ($\mu_l/\mu_s$) are not quantitatively comparable to some of the recent experimental data based on electron magnetic circular dichroism (EMCD), which have reported higher values for this ratio (0.08$\pm$0.01 for bcc Fe in Ref.~\cite{emcd1} and 0.14$\pm$0.03 for hcp-Co in Ref.~\cite{emcd2}). However, we found a closer agreement between theoretical results and the experimental data based on X-ray magnetic circular dichroism (XMCD) as shown in Table ~\cite{stearn,reck,chen}.

\section{Conclusions}
\label{conclusions}

In this work we have investigated the interatomic exchange of bcc Fe, hcp Co and fcc Ni, as one comes closer to the surfaces from the bulk region. Our theoretical method is based both on the local spin-density approximation, as well as dynamical mean field theory, in which dynamic correlations are treated explicitly. We have used a slab geometry in these studies, and found that for the central layers of the slabs, bulk like moments and exchange parameters are found in all three studied cases. As one approaches the surface region from the bulk, will observe a general trend of enhanced spin and orbital moments, both in LSDA and in DMFT. In fact, the difference between results of LSDA and DMFT is rather minor for these systems, at least when it comes to spin moments and interatomic exchange parameters. For the orbital moments we observe somewhat larger differences between LSDA and DMFT results, and we find that the latter compare in general better to experiments, where a comparison is possible. 

Inspection in more detail of the interatomic exchange interactions reveal a general trend of enhanced values at the surfaces. We find that this is primarily driven by the increased exchange splitting of the surface states, something which is caused by the reduced coordination number of surface atoms. Hence, the experimental observation of lower ordering temperatures of surfaces, which is a general phenomenon, is not caused by a reduction of the interatomic exchange interactions of the surfaces. In contrast the surface exchange interactions are enhanced. However, when coupled to an effective spin-Hamiltonian, of Heisenberg type or similar, the reduced coordination of surface atoms reduced the local Weiss field of the surface atoms, which makes them more susceptible to thermal fluctuations. 

Finally we have analyzed symmetry resolved aspects of nearest neighbor interactions of surface atoms of bcc Fe, and found that some of these interactions are ferromagnetic whereas some are antiferromagnetic, and that summed over all symmetry components, the nearest neighbor exchange interaction of surface atoms is antiferromagnetic. The magnetic order of the Fe surface is nevertheless ferromagnetic, due to strong ferromagnetic coupling to subsurface atoms. We argue however that the antiferromagnetic surface interactions of bcc Fe are inherent, and should be an avenue to tune complex magnetic structures of mono-atomic overlayers of Fe on bcc substrates.

\section*{Acknowledgement}
The authors thank J. Kudrnovsk\'y (FZU, Prague) for stimulating discussions and sharing his results prior to the publication. The computer simulations are performed on computational resources provided by NSC and UPPMAX allocated by the Swedish National Infrastructure for Computing (SNIC). We acknowledge financial support from the Swedish Research Council (VR), Energimyndigheten (STEM), the Knut and Alice Wallenberg Foundation, the Swedish e-Science Research Centre (SeRC), eSSENCE and the Swedish Foundation for Strategic Research (SSF). MIK acknowledges financial support by ERC grant No. 338957 and by NWO via Spinoza Prize. 

\appendix
\counterwithin{figure}{section}
\counterwithin{table}{section}

\section{The effect of U on J$_{ij}$}
\label{u}
\begin{figure}[th]  
 \includegraphics[width=0.43\textwidth]{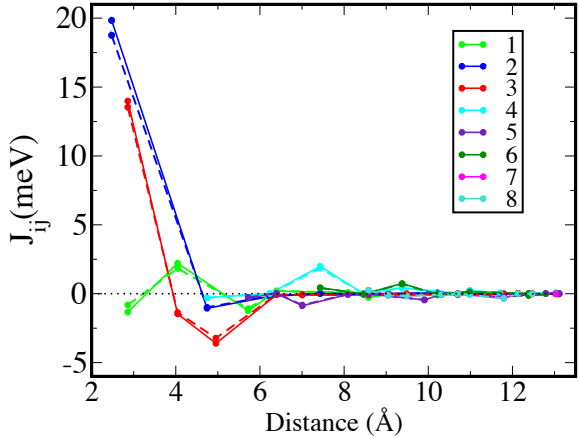}
 \caption{Layer-resolved exchange parameters ($J_{ij}$) for the surface atom of bcc Fe (atom \textit{i} is located at layer 1). The solid lines indicate the results obtained for U = 2.3 eV for all atoms, the dashed lines represent the results for layer-dependent U values (see the text). The layer numbering starts from surface denoted by 1, subsurface is denoted by 2 and so on.}
\label{fig:fe_u}
\end{figure}

At the surface, because of the less effective screening, the value of the Hubbard U is expected to increase. 
In this regard, we performed series of calculations using higher values for the surface and subsurface atoms to see the impact of U on the exchange parameters. For Fe, we have used U = 3 eV for atoms at the surface and U = 2.8 eV for those in the subsurface, while value for the inner layers is kept fixed to 2.3 eV. The outcome of these calculations, together with the results obtained with a uniform U value (2.3 eV) for all atoms, are shown in Fig.~\ref{fig:fe_u}. It is evident that larger U values for surface atoms result in a small uniform reduction of the parameters $J_{ij}$, but the trends are unchanged (comparison of the dashed lines with the solid lines of Fig.~\ref{fig:fe_u}). From the extent of these changes, we conclude that the overall behavior and more in particular the sign of the coupling do not change if U is varied within a reasonable range. Finally, similar conclusions can be obtained from analogous calculations for Co and Ni, which confirms that the values of the exchange parameters are rather robust with respect to the choice of the U value.

\section{Charge self-consistency}
\label{csc}

\begin{figure}[tp]
 \includegraphics[width=0.43\textwidth]{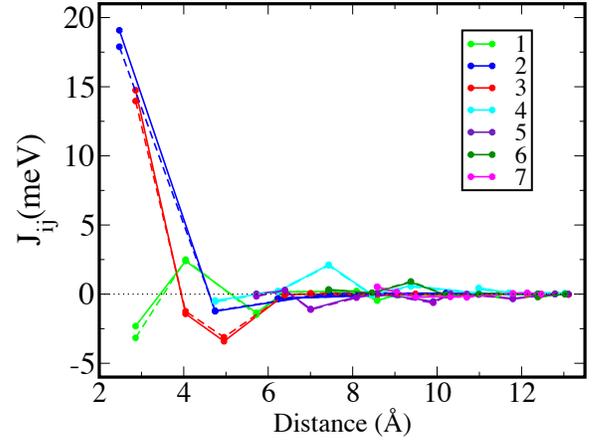}
 \caption{Layer-resolved exchange parameters ($J_{ij}$) for the surface atom of 7-layer Fe slab (atom \textit{i} is located at layer 1). The solid (dashed) lines indicate the results without (with) updating the electron density. The layer numbering starts from surface denoted by 1, subsurface denoted by 2 and so on.}
\label{fig:fe_csc}
\end{figure}

The results presented in the main text refer to calculations where the electron density is kept fixed to its LDA value, and the local correlation effects affect the results only through the self-energy function. We however performed several calculations to analyze the role of complete self-consistency over the electron density. We illustrate these results for a 7-layer slab of Fe(001). The obtained exchange parameters for one atom at the surface are reported in Fig. \ref{fig:fe_csc}, with and without updating the electron density. One can observe only small variations in the absolute magnitude of the parameters $J_{ij}$. These differences amount to only a few percent. Similar calculations have been performed for 7-layer slabs of Co and Ni, and lead to similar results. Thus, we conclude that the effects of charge self-consistence within the LDA+DMFT scheme are negligible for treating magnetic properties of the transition metal surfaces. This may be important for future investigations of thin films deposited on a substrate, where computational efficiency is going to be of primary importance.

\section{}
\label{qw}

In this appendix we show the results for the calculated orbital-resolved renormalisation factors $Z$ in slabs of Fe, Co and Ni.
$Z^{\sigma}_m$ denotes the inverse of an effective mass enhancement for the correlated orbital $m$ and is defined as:
\begin{eqnarray}
Z^{\sigma}_m=\biggl(1-\frac{d \text{Re} \Sigma^{\sigma}_{mm}(\omega)}{d\omega}\biggl|_{\omega=0}\biggl)^{-1},
\end{eqnarray}
where $\Sigma^{\sigma}_{mm}(\omega)$ is the self-energy projected on the orbital $m$ with the spin projection $\sigma$, $\omega=0$ corresponds to $E_F$.
In general, $Z$-factors are good measures of the strength of the correlation effects.
To have a compact description of the latter, we have also calculated the average $Z^{\sigma}_{avg}$ per spin channel.
$Z^{\sigma}_{avg}$ was computed using the following expression:
\begin{equation}
\label{eq:z}
Z^{\sigma}_{avg}= \frac{\sum_m Z^{\sigma}_m N_m(E_F)}{\sum_m N_m(E_F)}  ;  \qquad m =d_{z^2},...,d_{xy}
\end{equation}
where $N^{\sigma}_m(E_F)$ denotes the partial density of state at the Fermi level of a particular state. Thus, the average $Z$-factor is a weighted sum of orbital-resolved renormalisation factors.
The weight of each orbital is defined by its relative contribution to the spectral weight at the Fermi level.

\begin{table*}[thp]
\caption{\label{tab:table3} Orbital-resolved and average (avg.) renormalization factors $Z$ for Fe, Co and Ni slabs.}
\begin{tabular}{ |l||m{0.8cm}|m{0.9cm}|m{0.8cm}|m{0.8cm}|m{0.8cm}|m{0.8cm}|m{0.8cm}|m{0.9cm}|m{0.8cm}|m{0.8cm}|m{0.8cm}|m{0.8cm}| }
 \hline
 \multicolumn{13}{|c|}{Fe} \\
 \hline 
 & \multicolumn{6}{c|}{Majority Spin} &\multicolumn{6}{c|}{Minority Spin} \\ [1ex]
 \hline
 &\textrm{$d_{z^2}$} & \textrm{$d_{x^2-y^2}$}\hspace{-0.2cm} & \textrm{$d_{yz}$} & \textrm{$d_{xz}$} & \textrm{$d_{xy}$}& \textrm{avr}& \textrm{$d_{z^2}$} & \textrm{$d_{x^2-y^2}$}\hspace{-0.2cm} & \textrm{$d_{yz}$} & \textrm{$d_{xz}$} & \textrm{$d_{xy}$}& \textrm{avr} \\[1ex]
 \hline
 Surface  & 0.57&0.54&0.51&0.51&0.54&0.53&0.76&0.72&0.74&0.74&0.72&0.73\\[1ex]
Subsurface&0.73&0.72&0.72&0.72&0.72&0.72&0.78&0.77&0.81&0.81&0.81&0.80  \\[1ex]
 Middle &0.72&0.72&0.71&0.71&0.71&0.71&0.76&0.76&0.76&0.76&0.76&0.76\\
 \hline
\end{tabular}
\\[3ex]
\begin{tabular}{ |l||m{0.8cm}|m{0.9cm}|m{0.8cm}|m{0.8cm}|m{0.8cm}|m{0.8cm}|m{0.8cm}|m{0.9cm}|m{0.8cm}|m{0.8cm}|m{0.8cm}|m{0.8cm}| }
 \hline
 \multicolumn{13}{|c|}{Co} \\
 \hline
 & \multicolumn{6}{c|}{Majority Spin} &\multicolumn{6}{c|}{Minority Spin} \\ [1ex]
 \hline
 &\textrm{$d_{z^2}$} & \textrm{$d_{x^2-y^2}$}\hspace{-0.2cm} & \textrm{$d_{yz}$} & \textrm{$d_{xz}$} & \textrm{$d_{xy}$}& \textrm{avr}& \textrm{$d_{z^2}$} & \textrm{$d_{x^2-y^2}$}\hspace{-0.2cm} & \textrm{$d_{yz}$} & \textrm{$d_{xz}$} & \textrm{$d_{xy}$}& \textrm{avr} \\[1ex]
 \hline 
 Surface  &   0.73 & 0.75 &0.74&0.74&0.75&0.74&0.73&0.78&0.75&0.75&0.78&0.76 \\[1ex]
Subsurface&   0.78 & 0.77 &0.77&0.77&0.77&0.77&0.82&0.81&0.82&0.82&0.81&0.81\\[1ex]
 Middle &   0.78 & 0.77 & 0.76 &0.76&0.77&0.77&0.82&0.81&0.81&0.81&0.81&0.81\\
 \hline
\end{tabular}
\\[3ex]
\begin{tabular}{ |l||m{0.8cm}|m{0.9cm}|m{0.8cm}|m{0.8cm}|m{0.8cm}|m{0.8cm}|m{0.8cm}|m{0.9cm}|m{0.8cm}|m{0.8cm}|m{0.8cm}|m{0.8cm}| }
 \hline
 \multicolumn{13}{|c|}{Ni} \\
 \hline
 & \multicolumn{6}{c|}{Majority Spin} &\multicolumn{6}{c|}{Minority Spin} \\ [1ex]
 \hline
 &\textrm{$d_{z^2}$} & \textrm{$d_{x^2-y^2}$}\hspace{-0.2cm} & \textrm{$d_{yz}$} & \textrm{$d_{xz}$} & \textrm{$d_{xy}$}& \textrm{avr}& \textrm{$d_{z^2}$} & \textrm{$d_{x^2-y^2}$}\hspace{-0.2cm} & \textrm{$d_{yz}$} & \textrm{$d_{xz}$} & \textrm{$d_{xy}$}& \textrm{avr} \\[1ex]
 \hline 
 Surface  &0.78&0.77&0.79&0.79&0.78&0.78& 0.69&0.69&0.68&0.68&0.71&0.69\\[1ex]
Subsurface&0.82&0.82&0.80&0.80&0.81&0.81&0.77&0.75&0.75&0.75&0.75& 0.75\\[1ex]
 Middle &0.81&0.81&0.80&0.80&0.80&0.81&0.76&0.76&0.74&0.74&0.74&0.75\\
 \hline
\end{tabular}
\end{table*}

Calculated values of orbital-resolved $Z$-factors as well as their average values are shown in Table~\ref{tab:table3}.
An inspection of the results suggests that overall the correlation effects in all studied systems are not very strong (as a limiting case, $Z$=1 corresponds to LDA).
In Co and Ni slabs the renormalisation factors for all $3d$ orbitals are similar and their values lie in the range between 0.7 and 0.8.
It is also seen that the many-body effects are the most pronounced for surface electrons.
The largest renormalisation effects are found for Fe slab. 
In particular, majority-spin electrons of the surface atoms experience an almost twice mass enhancement.
Their overall spectral weight at the $E_F$ is relatively small (see. Fig. 3) and thus these quasiparticles are more sensitive to the addition of the self-energy.

Here we emphasize again the fact the differences in the $Z$-factors for majority and minority electrons are quite substantial.
However, within a particular spin channel the orbital-resolved $Z$-factors show relatively smaller deviations from the average value.

\bibliographystyle{apsrev4-1}

\providecommand{\noopsort}[1]{}\providecommand{\singleletter}[1]{#1}%
\end{document}